\documentclass[aps,prx,twocolumn,floatfix,superscriptaddress,showpacs,longbibliography,10pt]{revtex4-1}
\pdfoutput=1 
\usepackage[english]{babel}
\usepackage{amsmath,amssymb}
\usepackage[utf8]{inputenc}
\usepackage{graphicx}
\usepackage{color}
\usepackage{bm}
\usepackage{amssymb}
\usepackage{float}

\newcommand{\figref}[1]{Fig.~\ref{#1}}
\newcommand \Istvan[1]{\bgroup\noindent \textcolor{red}{#1}
\egroup\ignorespacesafterend}

\begin{document}
\title{Deterministic and stochastic models of dislocation patterning}

\author{Ronghai Wu}
\affiliation{Institute for Materials Simulation, Department of Materials Science,
Friedrich-Alexander University Erlangen-N\"urnberg (FAU), Dr.-Mack-Str. 77, 90762
F\"urth, Germany}

\author{Daniel Tüzes}
\affiliation{Department of Materials Physics, ELTE, Eötvös Loránd University,
H-1517 Budapest POB 32, Hungary}

\author{Péter Dusán Ispánovity}
\affiliation{Department of Materials Physics, ELTE, Eötvös Loránd University,
H-1517 Budapest POB 32, Hungary}

\author{István Groma}
\affiliation{Department of Materials Physics, ELTE, Eötvös Loránd University,
H-1517 Budapest POB 32, Hungary}

\author{Michael Zaiser}
\email{michael.zaiser@fau.de}
\affiliation{Institute for Materials Simulation, Department of Materials Science,
Friedrich-Alexander University Erlangen-N\"urnberg (FAU), Dr.-Mack-Str. 77, 90762
F\"urth, Germany}
\affiliation{Department of Mechanics and Engineering, Southwest Jiaotong University,
Chengdu, P.R. China}

\begin{abstract}
We study a continuum model of dislocation transport in order to investigate the formation of 
heterogeneous dislocation patterns. We propose a physical mechanism which relates the formation of 
heterogeneous patterns to the dynamics of a driven system which tries to minimize an internal energy 
functional while subject to dynamic constraints and state dependent friction. This leads us to a 
novel interpretation which resolves the old 'energetic vs. dynamic' controversy regarding the 
physical origin of dislocation patterns. We demonstrate the robustness of the developed patterning 
scenario by considering the simplest possible case (plane strain, single slip) yet implementing the 
dynamics of the dislocation density evolution in two very different manners, namely (i) a 
hydrodynamic formulation which considers transport equations that are continuous in space and time 
while assuming a linear stress dependency of dislocation motion, and (ii) a stochastic cellular 
automaton implementation which assumes spatially and temporally discrete transport of discrete 
'packets' of dislocation density which move according to an extremal dynamics. Despite the huge 
differences between both kinds of models, we find that the emergent patterns are mutually consistent 
and in agreement with the prediction of a linear stability analysis of the continuum model. We also 
show how different types of initial conditions lead to different intermediate evolution scenarios 
which, however, do not affect the properties of the fully developed patterns.
\end{abstract}

\maketitle

\section{Introduction}
Ever since the first TEM observations of dislocations it is known that the arrangement of 
dislocations in deformed crystals is practically never homogeneous: dislocations show an intrinsic 
propensity to forming heterogeneous patterns. There is an equally long standing discussion regarding 
the physical nature of these patterns which is matched by an amazing variety of approaches to their 
modeling, many of which are based upon analogies with pattern formation in other physical systems. 
Thus, it has been argued that dislocation patterns can be understood as minimizers of some kind of 
(elastic) energy functional (see e.g. \citet{Hansen1986_MSE}). Unfortunately this 'energetic' 
approach to dislocation patterning has hardly ever been cast into a tangible mathematical framework 
by actually formulating and minimizing the energy functional in question, one notable exception 
being the work of \citet{Holt1970_JAP} which is clearly crafted in analogy with contemporary models 
of spinodal decomposition patterns and relates the patterning of dislocation densities to the 
minimization of an associated internal energy functional -- a process which in stark contrast with 
experiment is predicted to occur even in the absence of external stress. The 'energetic' approach 
may be contrasted with the idea that dislocations in a deforming crystal constitute a driven system 
far from equilibrium where patterns may form as dissipative structures. This has led to the 
formulation of nonlinear sets of partial differential equations for dislocation densities 
\citet{Aifantis1985_JAP,Pontes2006_IJP} which give rise to a variety of interesting patterns. Some 
of these resemble dislocation patterns observed by TEM, while other types of patterns predicted by 
the same equations., such as spiral waves, have never been observed \cite{Salazar1995_AM}. In our 
opinion most of these models suffer from an overly phenomenological approach to modelling: They aim 
at reproducing patterns (actually, pictures of patterns) rather than deriving them from the known 
elastic and kinematic properties of dislocation systems. As a consequence, the fundamental 
controversy whether dislocation patterning is in essence an energetic or a dynamic phenomenon, which 
has been neatly summarized by \citet{Nabarro2000_PMA}, remains unresolved. 

Only in recent years, attempts have been made to formulate dislocation density based models based 
upon averaging procedures which lead from the dynamics of discrete dislocations to the evolution of 
dislocation densities in a systematic manner. These methods have been used to formulate the 
kinematics of dislocations in 2D and lately in 3D 
\cite{Hochrainer2007_PM,Hochrainer2014_JMPS,Hochrainer2015_PM} and also to systematically derive 
driving forces for the dynamics. To the latter end, two alternative approaches have been pursued: 
Driving forces for dislocation density evolution may be obtained by directly averaging the 
interaction forces of discrete dislocations \cite{Groma2003_AM,Valdenaire2016_PRB,Groma2016_PRB}. 
Alternatively, one may formulate an energy functional governing the dynamics based upon 
phenomenological considerations \cite{Groma2007_PM,Groma2015_PRL} or from direct averaging of the 
elastic energy of the discrete dislocation system \cite{Zaiser2015_PRB}, and then obtain driving 
forces from variation of the energy functional in conjunction with thermodynamic consistency 
requirements \cite{Hochrainer2016_JMPS}. Both approaches have been shown to yield mutually 
consistent results provided that, in the variational approach, a nontrivial mobility function is 
assumed which implements a friction stress \cite{Groma2016_PRB}.  

Importantly, these statistical averaging approaches do not pursue the primary aim of 'modelling 
patterning', i.e., of reproducing experimental images in a more or less faithful manner. Rather, 
such models aim at 'modelling dislocations': their primary thrust is to provide an adequate 
representation of the motion and interactions of dislocations in a continuum framework ('continuum 
dislocation dynamics', CDD). However, it has soon become clear that in CDD models the emergence of 
heterogeneous dislocation patterns turns out to be an almost inevitable feature of the collective 
dynamics. Simulations of CDD models in 3D demonstrate an intrinsic tendency towards dislocation 
patterning \cite{Sandfeld2015_MSMSE,Xia2015_MSMSE} as they relate the emergence of patterning to the 
same dislocation interactions that govern strain hardening, in line with the 'principle of 
similitude' \cite{Sauzay2011_PMS}. This principle has been related to fundamental invariance 
properties of the equations that govern the properties of discrete dislocation systems, and indeed 
all physically based models of dislocation patterning published in recent years are consistent with 
these invariance properties \cite{Zaiser2014_MSMSE}. 

While most recent models exploit advances in computational power \cite{Xia2015_MSMSE} or in 
kinematic averaging methods \cite{Sandfeld2015_MSMSE} in order to address the important problem of 
dislocation patterning in 3D and under conditions of multiple slip, the present authors have pursued 
a more simplistic yet more fundamental goal, namely to elucidate the respective influences of 
dynamic and energetic mechanisms on the patterning process. To this end we focus on a minimal system 
(2D, single slip) where an exact representation of the kinematics is possible and well-defined forms 
have been established both for the energy functional \cite{Groma2007_PM,Zaiser2015_PRB} and also for 
the effective mobility law \cite{Groma2016_PRB}. As a consequence, we can dispense to a large extent 
of phenomenological assumptions and obtain a complete understanding of the interplay of energy 
minimization, external driving, and friction in driving the emergence of dislocation patterns. In a 
previous work \cite{Groma2016_PRB} we had analyzed the linear stability of the ensuing equations. 
While this allowed to reach important conclusions regarding the patterning mechanism, a linear 
stability analysis is of necessity insufficient to decide upon the stability of the emergent 
patterns and the robustness of the patterning scenario. We therefore, in the present paper, complete 
this with a comprehensive study of nonlinear aspects of patterning including the influence of 
initial condition, pattern growth mode, and investigation of pattern stability as well as an 
investigation of the influences of model implementation (continuous vs. discrete, deterministic vs. 
stochastic). The paper is organized as follows: In Section 2, we briefly present the continuum model 
formulated by \citet{Groma2016_PRB} and its implementation in a spatially and temporally continuous 
setting. Section 3 presents a spatially and temporally discrete, stochastic model which considers 
the same spatial couplings and friction rules, but implements a completely different dynamics. 
Results from both models are presented in Section 4 and compared to results from linear stability 
analysis. A discussion concerning the nature and robustness of dislocation patterning is presented 
in the Conclusions. 

\section{Deterministic continuum model} 

In the following we give a brief summary of the continuum model of dislocation transport developed 
in \citet{Groma2016_PRB}, see also \citet{Valdenaire2016_PRB}. For a detailed discussion of the 
derivation and statistical averaging methodology the reader is referred to the original works of 
\citet{Groma2003_AM,Groma2016_PRB,Valdenaire2016_PRB}. We consider a 2D system of straight parallel 
edge dislocations of both signs which can be envisaged as charged point particles moving in a 
perpendicularly intersecting plane (taken to be the $xy$ plane). Dislocations of Burgers vector 
modulus $b$ are assumed to move on a single slip system which constrains their motion to the slip 
direction which we identify with the $x$ direction. 

\subsection{Transport equations}

The model equations have the structure of continuity equations. The stress-driven motion of a 
dislocation depends on its sign; we use the sign convention that, under a positive resolved shear 
stress, positive dislocations of density $\rho^+$ move with velocity $v^+$ in the $+x$ and negative 
dislocations of density $\rho^-$ move with velocity $-v^-$ in the $-x$ direction. These motions 
produce a shear strain $\gamma$ at rate
\begin{equation}
\partial_t \gamma = b \left(\rho^+ v^+ + \rho^- v^-\right) 
\end{equation}
Neglecting dislocation generation or annihilation, the transport equations have the simple structure
\begin{eqnarray}
 && \partial_t \rho^+(\bm{r},t) = - \partial_x (\rho^+ v^+)\nonumber\\
 && \partial_t \rho^-(\bm{r},t) = \partial_x (\rho^- v^-)
\label{eq:transport}
\end{eqnarray}
where
\begin{eqnarray}
 && v^+(\bm{r},t) = M_0 b {\cal T}^+(\bm{r},t) \nonumber\\
 && v^-(\bm{r},t) = M_0 b {\cal T}^-(\bm{r},t) \label{eq:velocities} .
\end{eqnarray}
In these equations, the ${\cal T}^{\pm}$ are effective shear stresses driving the motion of 
positive and negative dislocations and $M_0$ is a dislocation mobility coefficient (inverse drag 
coefficient). Hence, we assume the dislocation velocities to be proportional to the effective 
driving forces ${\cal T}^{\pm}b$ (i.e., the effective glide components of the Peach-Koehler forces.

\subsection{Evaluation of the effective driving stresses}

The effective driving stresses ${\cal T}(\tau_{\rm d}^{\pm},\tau_{\rm f}^{\pm})$ result from the 
combination of sign-dependent local driving stresses $\tau_{\rm d}^{\pm}$ and friction stresses 
$\tau_{\rm f}^{\pm}$ as
\begin{equation}
{\cal T}(\tau_{\rm d},\tau_{\rm f}) = \rm{sign}(\tau_{\rm d}) \left(|\tau_{\rm d}| - \tau_{\rm f}\right)
\label{eq:taueff}
\end{equation}
The driving stresses are given by combinations of a spatially homogeneous shear stress $\tau_{\rm 
ext}$ arising from remotely applied boundary tractions which provides the external driving force 
for dislocation motion and plastic flow, and a set of stress contributions describing dislocation 
interactions,
\begin{equation}
\tau_{\rm d}^{\pm} = \tau_{\rm ext} + \tau_{\rm int} + \tau_{\rm back} \pm \tau_{\rm diff}.
\label{eq:taudrive}
\end{equation}
We discuss the interaction stress contributions $\tau_{\rm int}$, $\tau_{\rm b}$, and $\tau_{\rm d}$ in turn:
\begin{enumerate}
\item
The 'internal' shear stress $\tau_{\rm int}$ arises from heterogeneity of the plastic eigenstrain
$\bm{\varepsilon}_{\rm pl} = \gamma (\bm{e}_x \otimes \bm{e}_y + \bm{e}_y \otimes \bm{e}_x)/2$. 
This stress can be calculated in various manners, e.g. by direct convolution of the dislocation 
shear stress field with the excess dislocation density $\kappa = \rho^+ - \rho^- = - \partial_x 
\gamma/b$ as done by \citet{Groma2003_AM}, by using an Airy stress function formalism 
\cite{Groma2007_PM}, or numerically by solving the eigenstrain problem using finite elements. In the 
following we adopt a fourth method where we calculate the internal stress directly from the plastic 
strain $\gamma$ using a Green's function formalism as adopted in \citet{Zaiser2005_JSTAT},
\begin{equation}
\tau_{\rm int}(\bm{r}) = \int \gamma(\bm{r'}) {\cal \mu}(\bm{r}-\bm{r'}) {\rm d}^2 
r'\label{eq:intstress}
\end{equation}
where ${\cal \mu}$ is an interaction kernel function with the Fourier transform
\begin{equation}
{\cal G}(\bm{k}) = \frac{\mu}{\pi(1-\nu)} \frac{k_x^2k_y^2}{k^4} = \mu T(\bm{k}), \label{eq:Gk}
\end{equation}
where $\mu$ is the shear modulus of the material, $\nu$ is Poisson's ratio, and $k_x$ and $k_y$ are 
components of the Fourier wavevector with modulus $k$. 
\item
The 'back stress' $\tau_{\rm back}$ counter-acts accumulation of dislocations of the same sign. It is given by  
\begin{eqnarray}
\tau_b(\bm{r})&=&-\mu b \frac{D}{\rho}\partial_x\kappa(\bm{r}) = \mu \frac{D}{\rho}\partial_{xx} 
\gamma(\bm{r}), \label{eq:backstress}
\end{eqnarray}
where $D$ is a nondimensional factor of the order of unity and $\rho = \rho^+ + \rho^-$ is the total 
dislocation density. We see that the back stress is proportional to the second gradient of the 
shear strain. 
\item
Finally, the 'diffusion stress' $\tau_{\rm diff}$ is given by
\begin{eqnarray}
\tau_{\rm diff}(\bm{r})&=&-\mu b \frac{A}{\rho}\partial_x\rho(\bm{r}),  \label{eq:diffstress}
\end{eqnarray}
where $A$ is another nondimensional factor of the order of unity. The terminology 'diffusion stress' 
is used because this stress, if inserted via Eqs (\ref{eq:taudrive}), (\ref{eq:taueff}), 
(\ref{eq:velocities}) into the transport equations Eq. (\ref{eq:transport}), gives rise to a 
diffusion-like contribution to the evolution of the total dislocation density $\rho$. 
\end{enumerate}
\citet{Groma2016_PRB} observed that the stress contributions $\tau_{\rm int}, \tau_{\rm back}$ and 
$\tau_{\rm diff}$ represent kinematic hardening contributions. Indeed all these stress contributions 
can, via variational calculus, be derived from an energy functional of the dislocation system given 
by
\begin{eqnarray}
&&E = E_{\rm el} + E_{\rm dis}, \nonumber\\
&&E_{\rm el} = \int \int \gamma({\bm r}){\cal G}(\bm{r}-\bm{r'})\gamma(\bm{r'}){\rm d}^2 r  {\rm d}^2 r',\nonumber\\
&&E_{\rm dis}= \int \mu b^2\left(A \rho \ln(\rho) + 
\frac{D}{2}\frac{\kappa^2}{\rho}\right){\rm d}^2 r. \label{eq:energy}
\end{eqnarray}
The first contribution to this functional represents the elastic energy associated with the average 
plastic eigenstrain $\gamma$, whereas the second term represents a correction which captures 
elastic energy contributions due to stress and strain heterogeneities on the scale of single 
dislocations, which cannot be represented in terms of the coarse grained strain variable $\gamma$. 
For a formal derivation of these terms by averaging the elastic energy of the underlying discrete 
dislocation system, see \citet{Zaiser2015_PRB}.

The 'friction stresses' $\tau_{\rm f}^{\pm}$ in the effective stress expressions (\ref{eq:taueff}) are given by
\begin{eqnarray}
\tau_{\rm f}^{\pm}  &=&\alpha \mu b \sqrt{\rho} \left(1 \mp \frac{\kappa}{\rho}\right). 
\label{eq:flowstress}
\end{eqnarray}
These stresses are of a different nature from the driving stresses: they represent friction-like, 
isotropic hardening contributions. While these stresses arise naturally from direct averaging of 
the dislocation interactions, they cannot be derived from an energy functional but need to be added 
'by hand' to an energy-based formalism where they enter in terms of a non-trivial, nonlinear 
mobility function with a mobility threshold \cite{Groma2016_PRB}. The functional form of these 
stresses is that of Taylor stresses; in physical terms, they represent the mutual trapping of 
positive and negative dislocations into dipolar or multipolar configurations. Their dependency on 
$\kappa$ reflects the fact that the presence of an excess of dislocations of one sign implies 
reduced pinning of the majority and enhanced pinning of the minority population. In particular, for 
$\kappa = \rho$ or $\kappa = - \rho$ (only positive or only negative dislocations) the pinning 
stress is zero. 

\subsection{Initial conditions, boundary conditions, loading protocol}

We implement periodic boundary conditions in $x$ and $y$ for the stresses, and in $x$ for the 
dislocation fluxes. For the stresses this means that the convolution integral in Eq. 
(\ref{eq:intstress}) is evaluated using the $L$-periodically continued Fourier transform of the 
kernel. As initial conditions we use
$\rho^{\pm}(\bm{r},t) = \rho_0/2 + \epsilon \delta\rho^{\pm}(\bm{r},t)$ where $\epsilon \ll 1$ and 
we consider two types of perturbation $\delta \rho^{\pm}$: (i) a Gaussian white noise of unit 
amplitude and (2) a localized Gaussian 'blob' of width $l = \rho_0^{-1/2}$ located at the center of 
the simulation cell. The loading protocol is simple: We impose a constant external stress and keep 
it fixed throughout the simulation, thus implementing creep-like testing conditions.

\section{Discrete stochastic model}

Our second model considers dislocation motion to be driven by the same stress contributions as 
introduced in Section 2B. However, the implementation of the dynamics differs radically. 

We now consider a spatially and temporally discrete model where space is discretized onto a $L\times 
L$ lattice  consisting of $N \times N$ square unit cells of size $d \times d$ with $d = L/N$. The 
simulation lattice is aligned with the $x$ and $y$ axes of a Cartesian coordinate system where, 
$(x,y) \to (i=x/d,j=yd)$. The discrete coordinate $i$ marks the slip direction and $j$ the direction 
of the slip plane normal. Periodic boundary conditions are assumed. Again we consider densities 
$\rho^{\pm}$ of positive and negative dislocations, however, dislocation densities are now assumed 
to be constant over each lattice cell and to be 'quantized' in units $\rho_{\rm d}^{\pm}$ which are 
integer fractions $\rho^{\pm}_{\rm d} = \rho_0/ (2M)$ of the overall dislocation density $\rho_0$. A 
discrete density quantum $\rho_{\rm d}$ of sign $s \in \{1,-1\}$ is henceforth denoted as a positive 
or negative 'dislocatom'. The dislocation state of lattice site $(i,j)$ is then characterized by the 
densities $\rho^+_{ij} = n_{ij}^+\rho_{\rm d}$ and $\rho^-_{ij} = n^-_{ij}\rho_{\rm d}$ or 
equivalently by the respective dislocatom numbers. Again, we consider the overall densities of 
positive and negative dislocations and hence the total dislocatom numbers to be conserved. 

We evolve the quantized dislocation densities on the lattice in discrete steps. Since all dependent 
and independent variables of the problem can be expressed in terms of integer numbers, we are 
dealing with a cellular automaton (CA) dynamics for which we now specify the evolution rules. 

\subsection{Cellular automaton dynamics}

The motion of dislocations is described as discrete shuffling of dislocatoms between sites that are 
adjacent in $i$ direction. Motion of positive and negative dislocatoms is controlled by driving 
forces which are proportional to the same effective stresses ${\cal T}^{\pm}$ that govern 
dislocation transport in the continuum model, with the only difference that these stresses (and also 
the plastic strains) are now defined on the boundaries between cells $(i,j)$ and $(i+1,j)$ that are 
adjacent in the slip direction. This implies that the lattice used for stress evaluation is shifted 
with respect to the lattice used for dislocation density evolution by $d/2$ along the direction of 
slip. Without loss of generality, we denote as boundary $(i,j)$ the boundary between cells $(i,j)$ 
and $(i+ 1,j)$. 

Dislocatoms move across boundaries subject to the following rules:
\begin{itemize}
\item
Dislocatoms do not move across boundaries experiencing zero effective stress. 
\item
Among all boundaries experiencing non-zero effective stress we determine, in each given time step, a 
critical boundary and dislocatom sign defined as the boundary and sign for which the effective 
stress has the largest absolute value:
\begin{equation} 
(i_{\rm m},j_{\rm m},s_{\rm m}): |{\cal T}^{s_{\rm m}}_{i_{\rm m}j_{\rm m}}| := \max_{i,j,s} |{\cal T}^{s}_{ij}|.
\end{equation}
Across this critical boundary, we move one dislocatom of sign $s_{\rm m}$ in direction $s_{\rm m} {\rm sign}({\cal T}^{s_{\rm m}}_{i_{\rm m}j_{\rm m}})$. In other words, positive dislocatoms move to the right from site $(i\to i+1)$ under a positive stress and to the left from $(i+1\to i)$ under a negative effective stress, while negative dislocatoms show the opposite behavior. After a dislocatom has moved across a boundary, the dislocatom numbers on both sites are adjusted accordingly.
\item
Motion of a dislocatom across a boundary changes the strain associated with this boundary. If a 
dislocatom moves from a site under a positive effective stress, then the strain $\gamma_{ij}$ on the 
crossed boundary is increased by $\rho_{\rm d} b d$. If the dislocatom moves under a negative 
effective stress, the strain is decreased by the same amount. 
\item 
After a dislocatom has moved we re-calculate all stresses (for details see below) and determine the next critical boundary.
\end{itemize}
These rules implement a CA with extremal dynamics, corresponding the a physical situation where the 
velocity of dislocations increases with effective stress in a very abrupt manner (e.g. an 
exponential law with a large exponent or a very high power law), such that the dislocation with the 
highest stress moves much faster than all others. In this sense the CA dynamics provides an extreme 
contrast with the linear velocity law assumed in the continuum model. 

\subsection{Calculation of stresses}

The effective driving stresses ${\cal T}^{\pm}_{ij}$ are calculated from the same equations as for 
the transport model, with some adjustments for the discrete nature of the model and for the 
inclusion of stochastic terms. The total and excess dislocation densities are evaluated as 
$\rho_{ij} = \rho_{\rm d}(n^+_{ij} +n^-_{ij})$ and  $\kappa_{ij} = \rho_{\rm d}(n^+_{ij} - 
n^-_{ij})$. The external stress is constant throughout the system. Internal stresses are evaluated 
according to Eq. (\ref{eq:intstress}) with the convolution replaced by the discrete lattice sum. 
Back stresses and diffusion stresses are evaluated from Eqs. (\ref{eq:backstress}) and 
(\ref{eq:diffstress})
with the spatial derivatives replaced by the respective directional difference quotients. The 
friction stress associated with the boundary $(i,j)$ is evaluated as
\begin{eqnarray}
\tau_{{\rm f},ij}^{\pm} &=&\alpha \mu b \sqrt{\rho_{ij} + \rho_{i+1,j}} \left(1 \mp 
\frac{\kappa_{ij}+\kappa_{i+1,j}}{\rho_{ij} + \rho_{i+1,j}}\right) * \xi_{ij}.\nonumber\\ 
\label{eq:flowstressfluct}
\end{eqnarray}
where $\xi_{ij}$ is a Gaussian distributed random variable of average $1$ and standard deviation 
$\sigma_{\tau}$. After a dislocatom move across some boundary, a new value of this variable is 
assigned to the boundary from the same distribution. This feature allows the model to account for 
stress fluctuations arising from the changes in local configurations of the discrete dislocations. 
Setting $\sigma_{\tau}=0$ makes Eq. (\ref{eq:flowstressfluct}) the direct discrete counterpart of 
(\ref{eq:flowstress}). 

\subsection{Initial conditions, boundary conditions, loading protocol}

We impose periodic boundary conditions as in the continuum model. Initial conditions are constructed
by placing $N\times N\times M/2$ positive and an equal number of negative dislocatoms randomly on 
the simulation lattice sites. We use two different types of loading protocol: (i) we impose a 
constant stress as in the continuum model. (ii) Alternatively, we increase, after an initial 
relaxation step, the stress precisely to the value needed to create one critical boundary. We trace 
the subsequent relaxation until no critical boundaries are left, and repeat. This algorithm 
corresponds to a quasi-static (infinitely slow) increase of the external stress and produces a 
stress-strain curve which approaches a horizontal asymptote corresponding to the macroscopic flow 
stress.  

\section{Results}

\subsection{Linear Stability Analysis of Transport Equations}

A linear stability analysis (LSA) of the dislocation transport equations in Section 2 has been performed 
by \citet{Groma2016_PRB}. Here we briefly summarize the results as a reference for comparison with 
the numerical investigation of the fully nonlinear equation and with the results obtained from the 
discrete stochastic model. One considers a spatially homogeneous reference state $\rho^+ = \rho^- = 
\rho_0/2$ under external stress $\tau_{\rm ext}$ and investigates the time evolution of 
infinitesimal perturbations around this state in linear approximation. Because of the general 
scaling invariance properties of dislocation systems, results can be expressed in a generic form 
where all dislocation densities are measured in units of $C_{\rho}=\rho_0$, all lengths in units of 
$C_l = \rho_0^{1/2}$, all times in units of $C_t = (M_0 \mu b^2 \rho_0)^{-1}$, all stresses in units 
of $C_{\tau} = \mu b \sqrt{\rho}$, and all strains in units of $C_{\gamma}=b\sqrt{\rho}$. In the 
following we use these units throughout. 

Linear stability analysis leads to the following results:
\begin{itemize}
\item 
No plastic flow occurs and the dislocation microstructure is static for $\tau_{\rm ext} \le \alpha$. 
\item
In the flowing phase ($\tau_{\rm ext} > \alpha$), the growth rates of fluctuations follow from the equation
\begin{eqnarray}
&&\Lambda^{\pm}(\bm{k}) = -\frac{(A+D)k_x^2+T(\bm{k})}{2} \nonumber\\
&&\pm\frac{\sqrt{[(D+A)k_x^2+T(\bm{k})]^2-4k_x^2[A(Dk_x^2+T(\bm{k}))-B]} }{ 2 }\nonumber\\
\label{eq:Lambda}
\end{eqnarray}
where 
\begin{equation}
B = \tau_{\rm ext}[(3/2) \alpha -\tau_{\rm ext}].
\end{equation}
\item
Within the unstable stress regime, there exists a band of unstable wave-vectors $\bm{k}$ fulfilling the equation
\begin{equation}
ADk_x^2 + AT(\bm{k})-B < 0.
\end{equation}
Perturbations of maximum amplification have the wave-vector $\bm{k}^{\rm max}$ with $k_y^{\rm max}=0$ and
\begin{equation}
 k_x^{\rm max}= \rho_0^{1/2} \left[2 B \frac{-1 + \sqrt{1 + \frac{(A-D)^2}{4AD}}}{(A-D)^2}\right]^{1/2}.
\label{eq:kmax}
\end{equation}
Hence, we expect the emerging patterns to be dominated by heterogeneities in $x$ rather than $y$ direction.
\end{itemize}
\begin{figure*}[t]\centering
\includegraphics[angle=0,width=17cm]{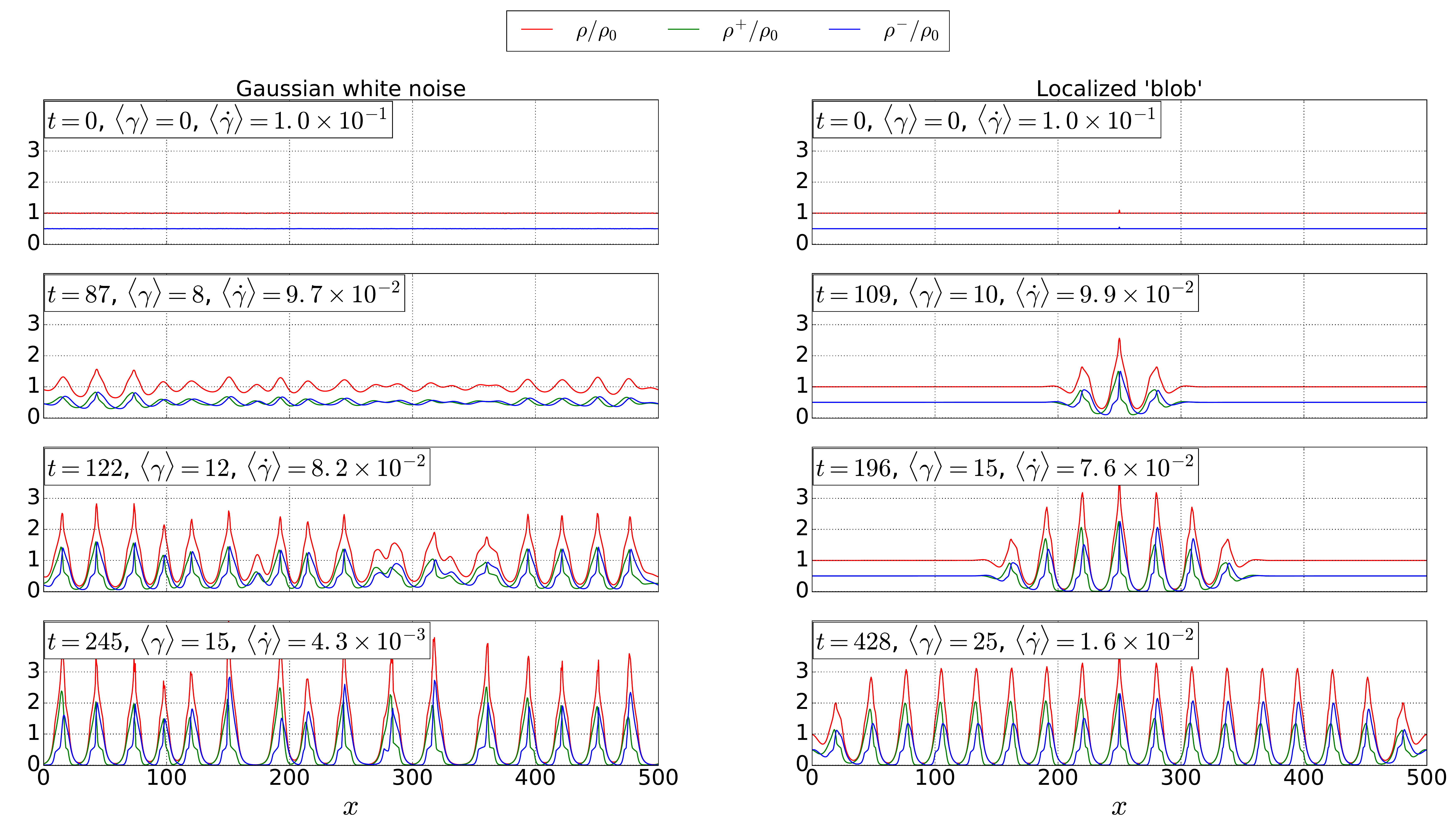}
\caption{\label{fig:Ronghai_1D_patterns}Spatio-temporal evolution of dislocation density patterns for two different initial conditions; left: small Gaussian white noise superimposed on a homogeneous density distribution, right: localized density fluctuation superimposed on a homogeneous distribution; 
parameters $A=0.5, D=0.4, \alpha = 0.3, \tau_{\rm ext}=1.1 \alpha$.}
\end{figure*}
We now compare
these predictions with solutions of the nonlinear transport equations, both for continuous transport and for
discrete stochastic 'dislocatom' dynamics.

\subsection{Simulations of the continuum transport equations}

\begin{figure}[htb]
\includegraphics[angle=0,width=9cm]{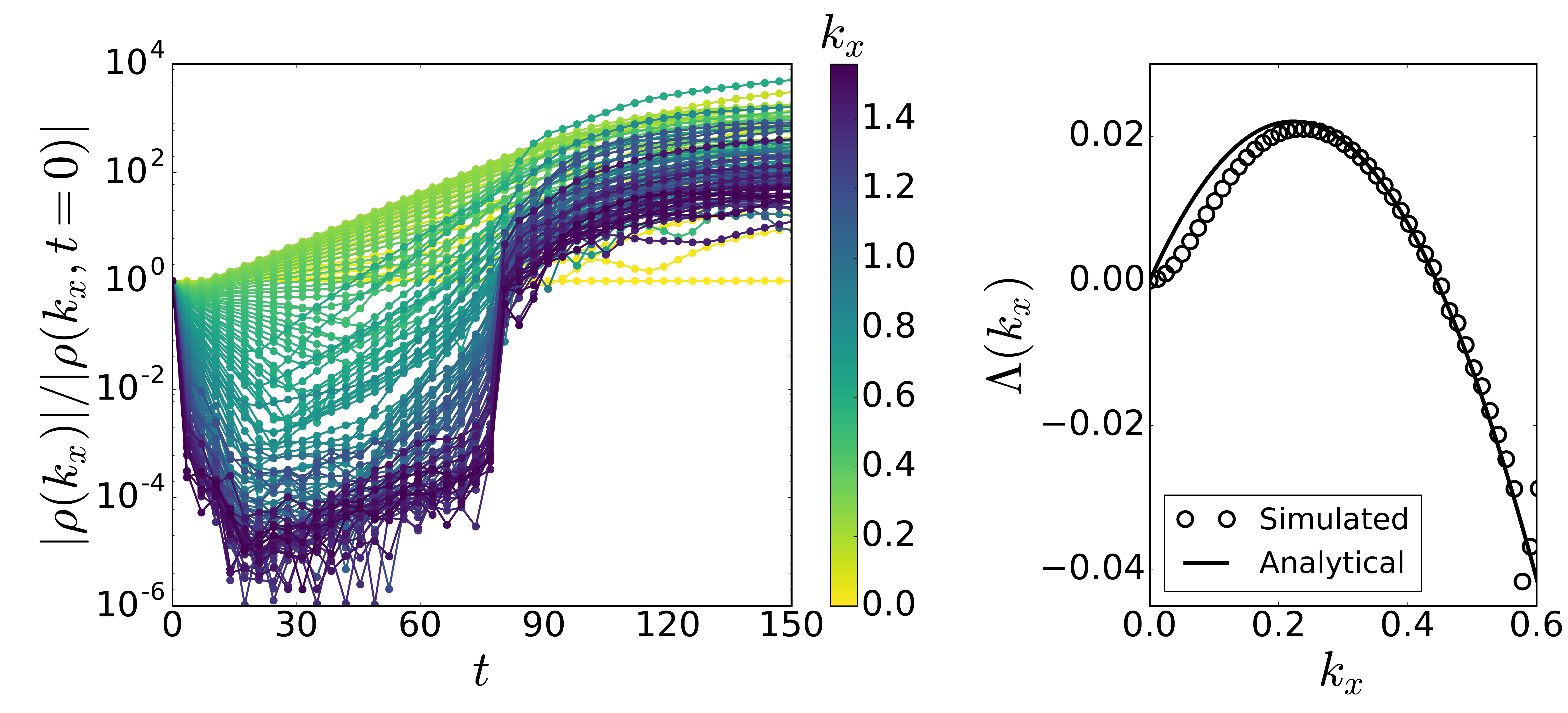}
\caption{\label{fig:Ronghai_1D_FFT}Left: Evolution of the Fourier modes $\rho(k_x)$ of the total dislocation density 
$\rho(x)$; right: growth rates as deduced from the initial slope of the $\ln |\rho(k_x)|$ vs $t$ 
curves and analytical prediction according to Eq.  (\ref{eq:Lambda}); parameters as in Figure 1.}
\end{figure}
\begin{figure}[htb]
\includegraphics[angle=0,width=9cm]{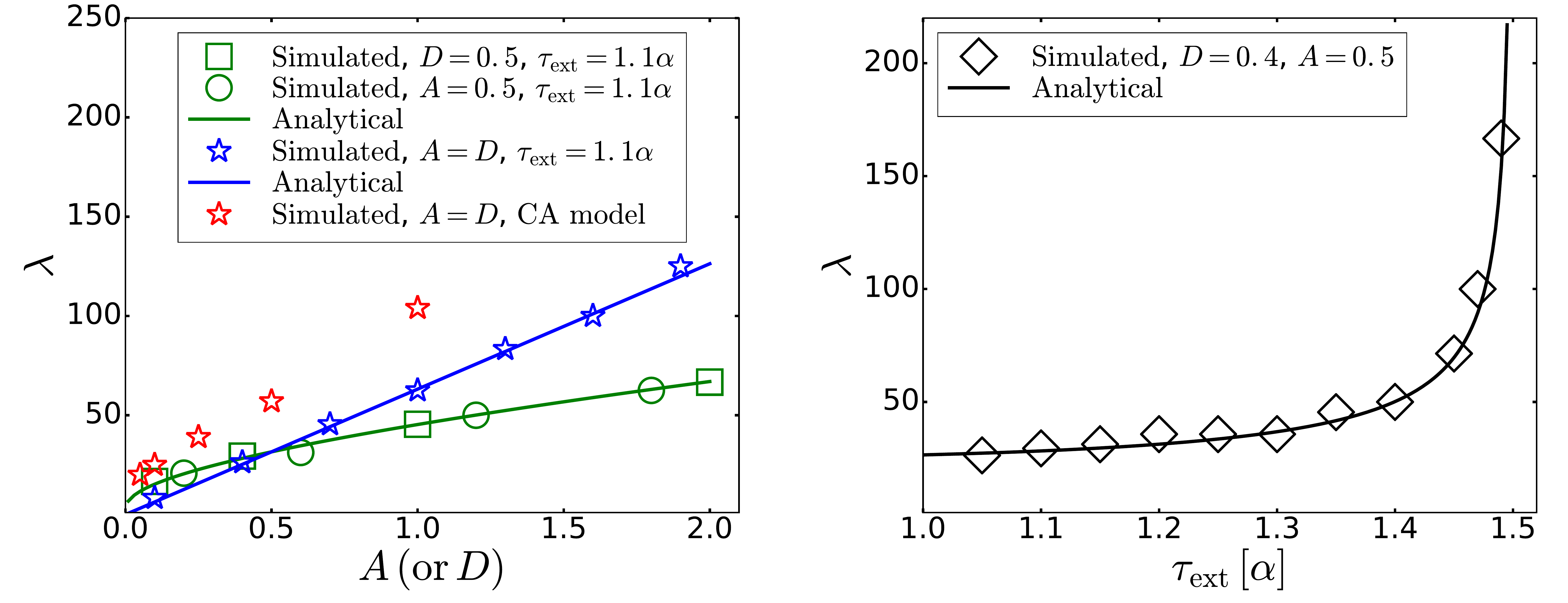}
\caption{\label{fig:wavelength}Pattern wavelength $\lambda$ as a function of the model parameters, left: $\lambda(A,D)$ values for $D=0.5$ (squares), $A=0.5$ (circles) and $A=D$ (stars), solid line: theoretical curve according to Eq. (\ref{eq:kmax}) (note that this expression is symmetrical with respect to an exchange of $A$ and $D$), $\tau_{\rm ext}=1.1 \alpha$; right: $\lambda(\tau_{\rm ext})$ for $A=0.5$ and $D=0.4$, solid line: theoretical curve according to Eq. (\ref{eq:kmax}). }
\end{figure}
Given that LSA predicts the dominant unstable mode to be associated with heterogeneities along the 
$x$ but not the $y$ direction, we first investigate a one-dimensional scenario where we impose 
homogeneity in $y$ direction, hence $k_y$ = 0 by construction. In $x$ direction we use periodic 
boundary conditions with period $L = 500$. We consider two types of initial conditions, namely (i) a 
Gaussian white noise and (ii) a small, localized dislocation density 'blob' on top of the 
homogeneous background (\figref{fig:Ronghai_1D_patterns}). The amplitudes of the Fourier components of the 
perturbation are identical in both cases, however, in case of the localized 'blob' the phases are 
identical whereas for the white noise they are random. Assuming a white noise perturbation leads to 
spatially distributed growth of the patterns, whereas a localized blob as initial condition leads to 
a correlated growth scenario where a fully developed pattern emerges locally and then spreads 
through propagation of an enveloping wave. Despite the different growth dynamics, the fully 
developed patterns resulting from initial conditions (i) and (ii) are very similar in terms of 
morphology and wavelength. The patterns consist of periodic walls of high dislocation density 
separated by dislocation depleted channels. The increased dislocation density in the walls implies a 
high flow stress leading to piling up of positive dislocations on the left and of negative 
dislocations on the right side of the walls. The ensuing back stresses, in turn, reduce dislocation 
flow in the intermediate channels: Dislocation patterning is associated with a reduction in the 
overall strain rate.

\begin{figure*}[p]
\includegraphics[angle=0,width=12cm]{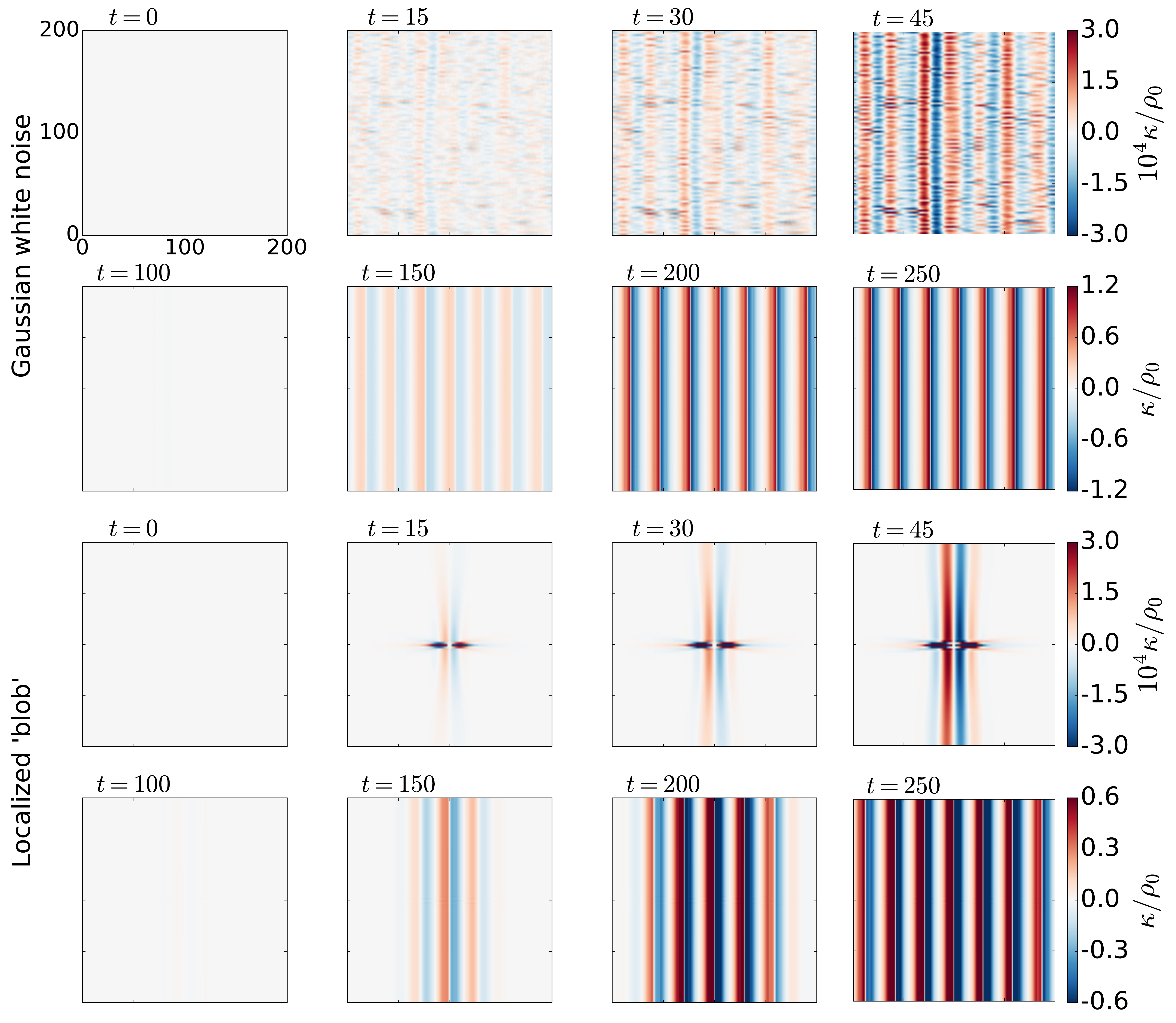}
\caption{\label{fig:Ronghai_2D_kappa}Spatio-temporal evolution of dislocation density patterns (excess density $\kappa(\bm{r})$) in two dimensions for two different initial conditions; upper two rows: small Gaussian white noise superimposed on a homogeneous density distribution; lower two rows: localized density fluctuation superimposed on a homogeneous distribution; parameters as in Figure 1.}
\includegraphics[angle=0,width=12cm]{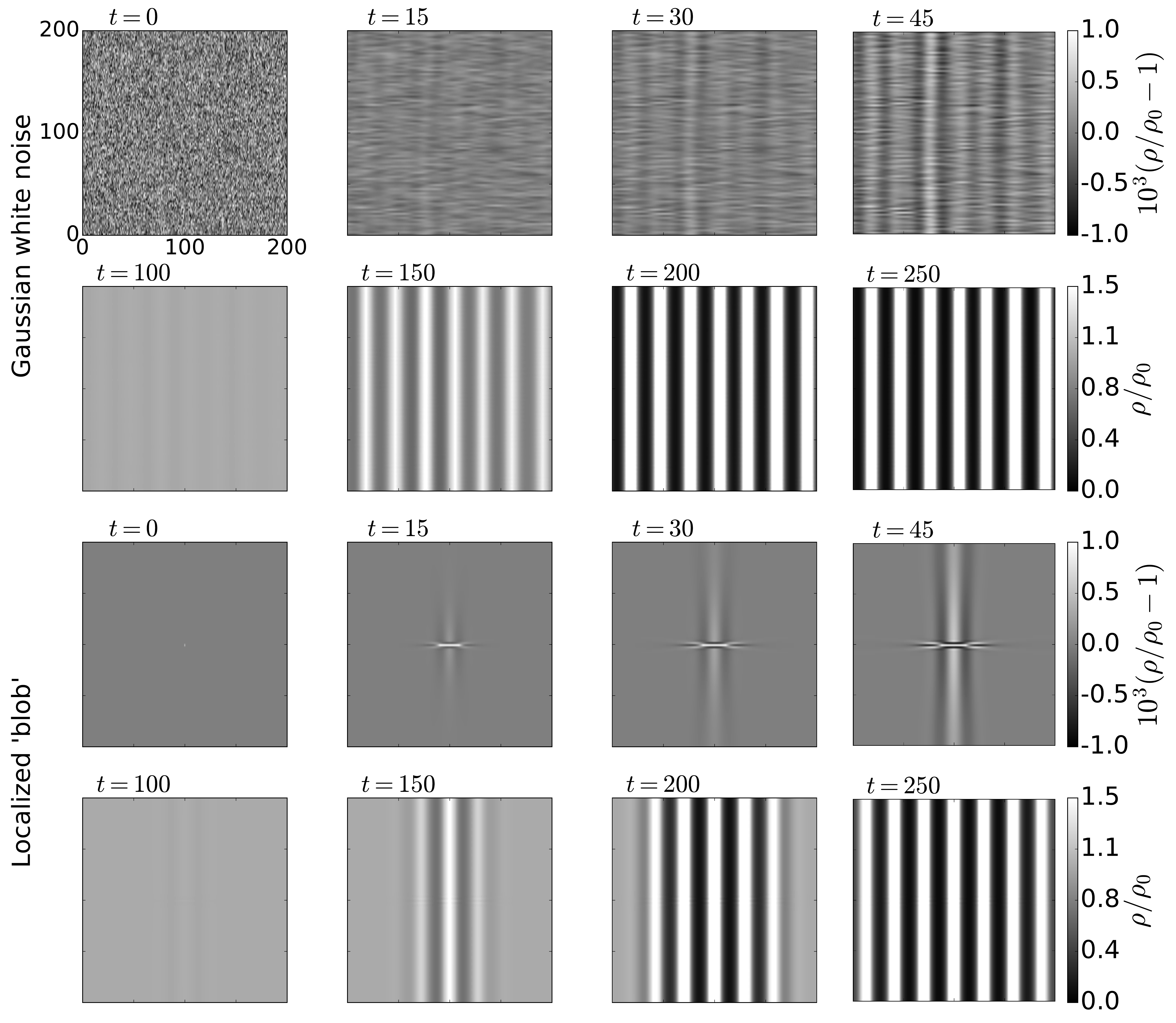}
\caption{\label{fig:Ronghai_2D_rho}Spatio-temporal evolution of dislocation density patterns (total density $\rho(\bm{r})$) in two dimensions for two different initial conditions; upper two rows: small Gaussian white noise superimposed on a homogeneous density distribution; lower two rows: localized density fluctuation superimposed on a homogeneous distribution; parameters as in Figure 1.}
\end{figure*}

Initially all Fourier modes of the perturbation have equal amplitude in both cases. The time 
evolution of the Fourier coefficients of the emergent patterns is shown in \figref{fig:Ronghai_1D_FFT} (left) 
for case (i); case (ii) shows a practically identical behavior. From the initial growth rates of 
the discrete Fourier modes $\rho(k_x)$ we deduce growth factors defined as $\Lambda(k_x) = \Delta 
\ln\rho(k_x)/\Delta t$. Comparison with the analytical prediction of Eq. \ref{eq:Lambda}  shows 
excellent agreement as illustrated in \figref{fig:Ronghai_1D_FFT} (right) . The wavelengths of the fully 
developed patterns match very closely (within 5\%) the predictions of linear stability analysis for 
the wavelength of the mode with maximum amplification. This observation, which holds throughout the 
parameter regime (\figref{fig:wavelength}), is remarkable since the nonlinearities have clearly a strong 
influence on the density distribution which is very different from a sinusoidal wave.  

We then study the same patterning scenarios in two dimensions. In this case the emergent patterns 
have a stripe-like character where the system is near-homogeneous in $y$ direction whereas the $x$ 
dependency of the dislocation densities is almost identical to the one-dimensional case. If we use a 
Gaussian white noise as initial perturbation, embryonic patterns start growing locally and then, in 
a first 'synchronization' stage organize in $y$ direction to form parallel walls. In a second 
'growth' stage the amplitude of these wall like dislocation density modulations increases while the 
once-established pattern remains in place (\figref{fig:Ronghai_2D_kappa} and \figref{fig:Ronghai_2D_rho}, top). If, on the other 
hand, we start from a localized dislocation density 'blob' then an interesting scenario occurs 
(\figref{fig:Ronghai_2D_kappa} and \figref{fig:Ronghai_2D_rho}, bottom): The blob causes positive and negative dislocations to pile up 
from both sides. The long range stresses of the double pile up then lead to growth of a double wall 
similar to a kink band in $y$ direction. Finally, the double wall serves as nucleus for a nonlinear 
wave which spreads the pattern in $y$ direction as in the one dimensional case. Irrespective of the 
growth mode, the wavelength and morphology of the patterns are almost identical to the one 
dimensional case. 

\begin{figure}[htb]
\includegraphics[angle=0,width=9cm]{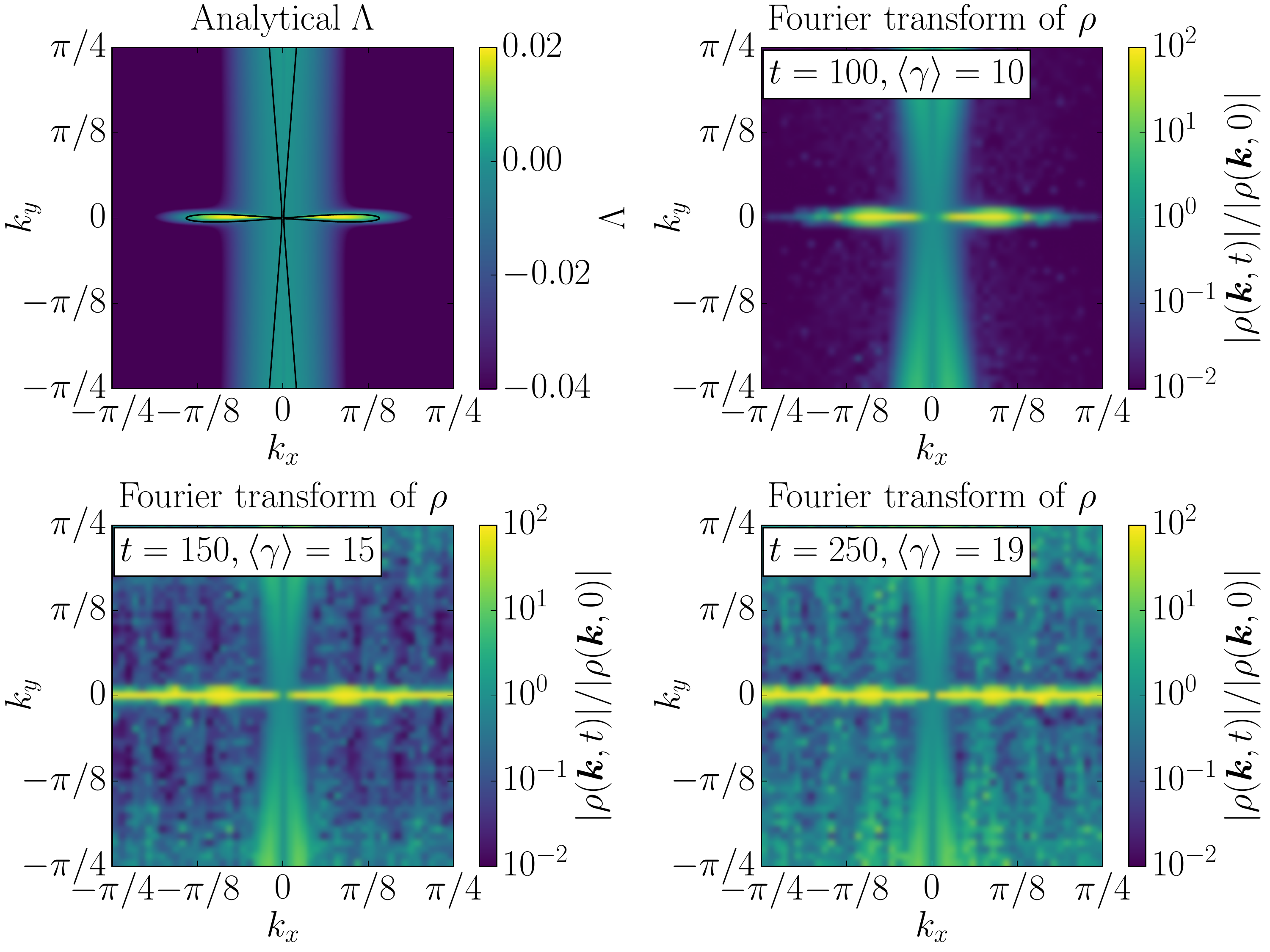}
\caption{\label{fig:Ronghai_2D_FFT}Top left: Analytical growth factor $\Lambda$ as in Figure 2; top right: Normalized Fourier 
pattern $\rho(\bm{k},t)$ for $t=100 C_t$, strain $\langle \gamma \rangle = 10$; bottom: Fourier 
patterns $\rho(\bm{k},t)$ for $t=150 C_t$, strain $\langle \gamma \rangle = 15$ and $t=250 C_t$, 
$\langle \gamma \rangle = 19$; parameters as in Figure 1.}
\end{figure}
Finally, \figref{fig:Ronghai_2D_FFT} shows the Fourier spectrum of the emerging dislocation density 
distribution. We use a logarithmic scale, hence the color level can also be envisaged as an 
exponential growth factor, enabling direct comparison with \figref{fig:Ronghai_2D_FFT}, top left. It can be 
seen that the Fourier pattern of the developing pattern closely matches the growth predictions of 
linear stability analysis also in 2D (\figref{fig:Ronghai_2D_FFT}, top right). At later stages, nonlinear 
effects lead to growth also of initially damped short-wavelength modes (\figref{fig:Ronghai_2D_FFT}, bottom). This is in close analogy with the 1D observations shown in \figref{fig:Ronghai_1D_FFT}, left. Note that the growth of damped modes concerns mainly harmonics in $x$ of the initial unstable 
mode, as evidenced by the periodic striations of the Fourier patterns in \figref{fig:Ronghai_2D_FFT}, bottom. 

\subsection{Simulations of the stochastic cellular automaton model}

\begin{figure}[htb]
\includegraphics[angle=0,width=6cm]{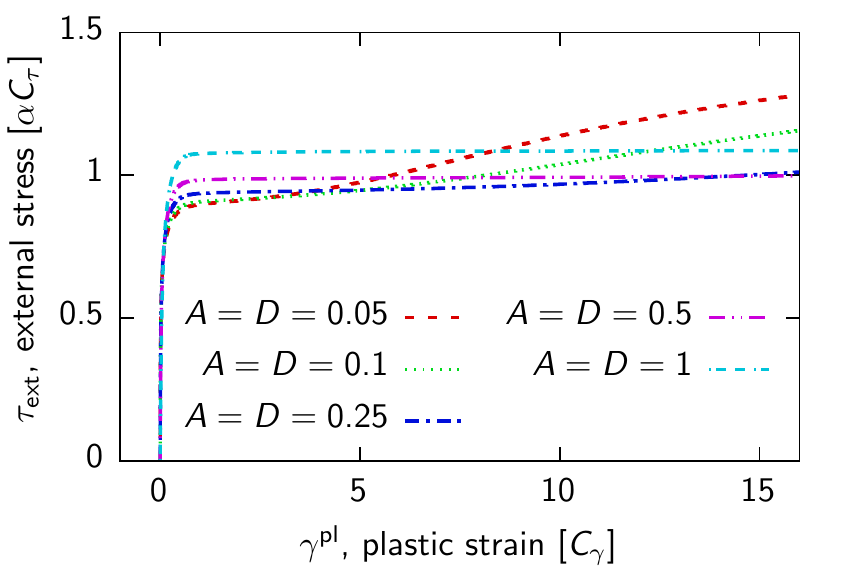}
\caption{\label{fig:Daniel_stress_strain_curve}Stress-strain curves as obtained from the stochastic CA model under quasistatic loading conditions.}
\end{figure}

\begin{figure}[htb]
\includegraphics[angle=0,width=8cm]{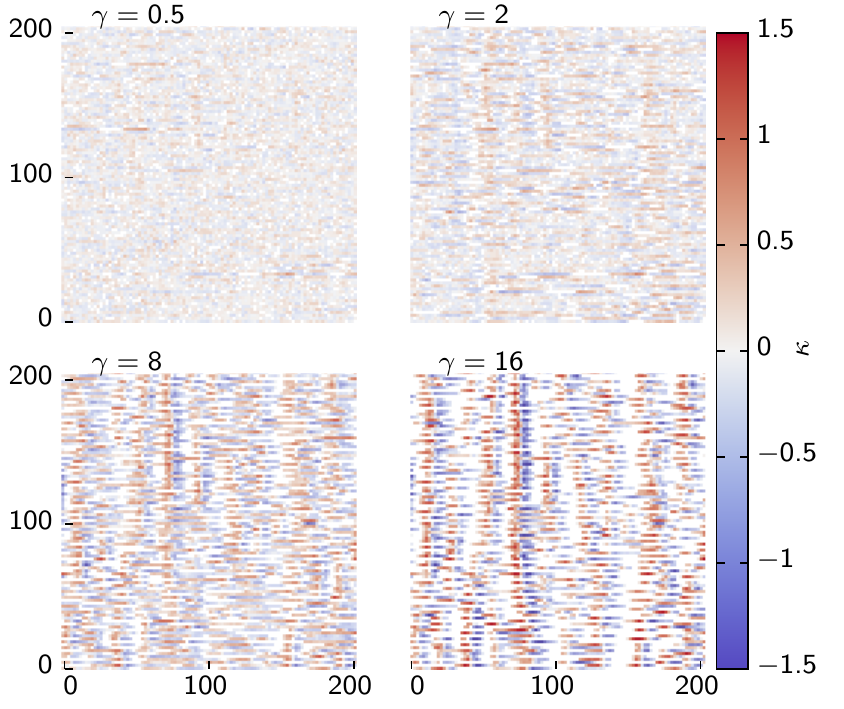}
\caption{\label{fig:CA_2D_kappa}Spatio-temporal evolution of dislocation density patterns (excess density $\kappa(\bm{r})$) 
as obtained from the stochastic CA model using a quasi-static loading protocol; parameters 
$A=D=0.1$, $\alpha = 0.3$.}
\end{figure}

Simulations of the stochastic cellular automaton model were conducted using a quasi-static stress 
increase protocol as described in Section IIIC with the dislocatom size corresponding to $M=16$. 
This leads to ascending stress strain curves  
reaching a horizontal asymptote (\figref{fig:Daniel_stress_strain_curve}). 
Pattern formation is illustrated in \figref{fig:CA_2D_kappa}. We can see the emergence of alternating 
walls of positive and negative dislocations which become more pronounced with increasing strain. 

\begin{figure}[htb]
\includegraphics[angle=0,width=4cm]{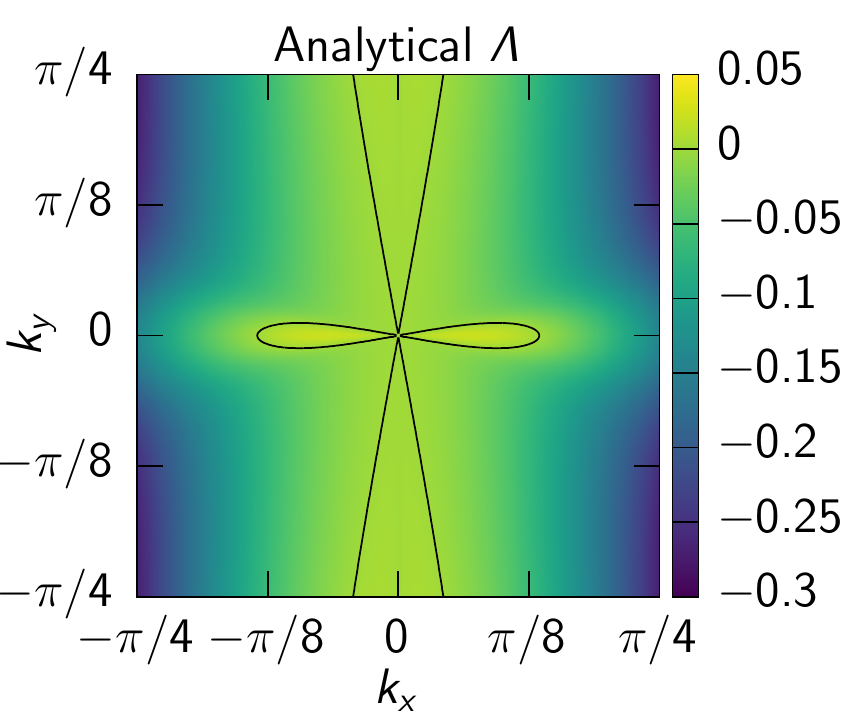}
\includegraphics[angle=0,width=4cm]{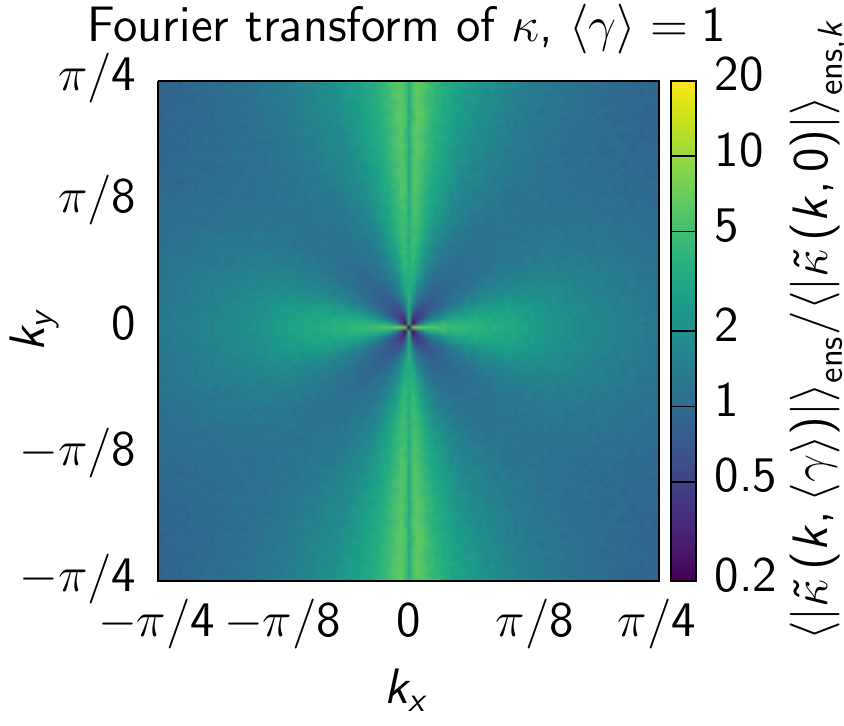}
\includegraphics[angle=0,width=4cm]{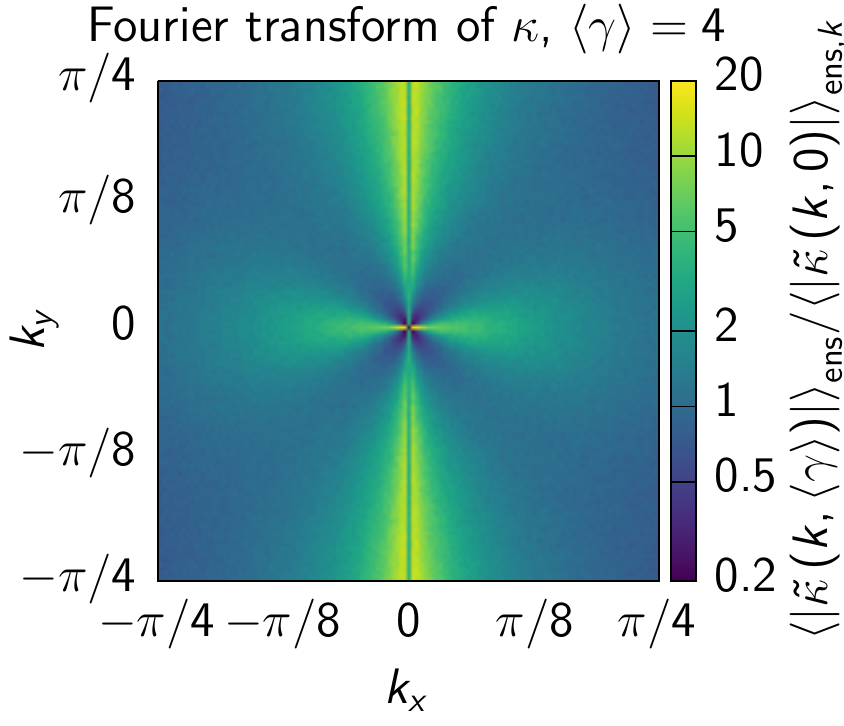}
\includegraphics[angle=0,width=4cm]{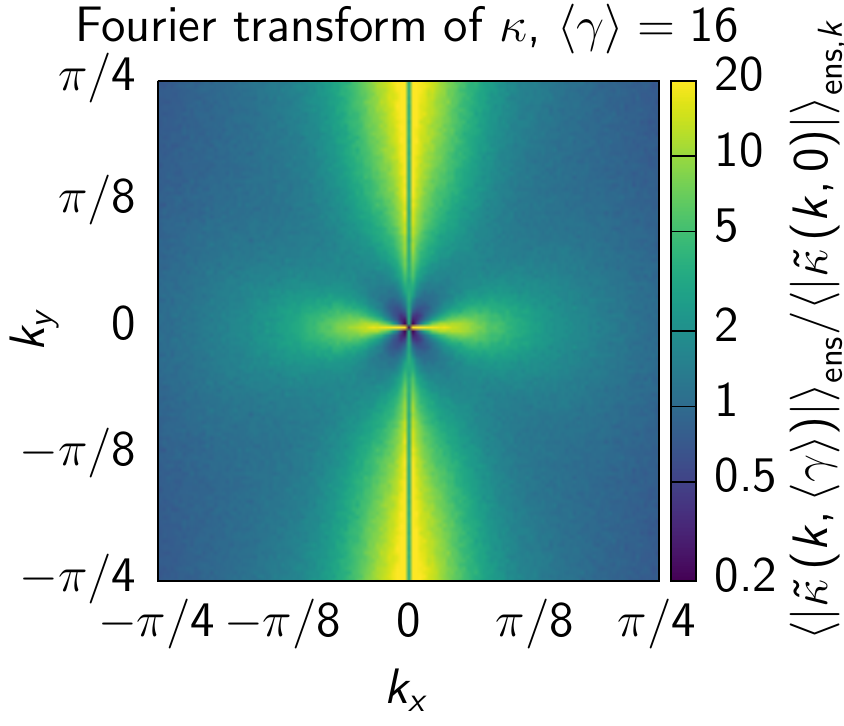}
\caption{\label{fig:Daniel_2D_FFT}Top left: Analytical growth factor $\Lambda$ according to Eq.(\ref{eq:Lambda}) at A=D=0.5; 
top right: Normalized Fourier 
patterns $\kappa(\bm{k},\gamma)$ for mean strain $\langle \gamma \rangle = 1$; bottom: Fourier 
patterns $\kappa(\bm{k},\gamma)$ for mean strains $\langle \gamma \rangle = 4$ and $\langle \gamma 
\rangle  = 16$, the Fourier patterns have been averaged over 100 realizations of the CA dynamics; 
cell size $d=2$.}
\end{figure}

The Fourier transform of the emergent patterns, taken at different strains, points to a growth 
scenario that differs substantially from that in the deterministic transport model. While the 
overall symmetry of the Fourier pattern matches the observations from the deterministic transport 
model and the corresponding linear stability analysis results, \figref{fig:Daniel_2D_FFT} demonstrates that 
the dominant wavelength of the patterns obtained from the CA shifts in the course of patterning 
from shorter to larger wavelengths (smaller $k_x$). This may be a feature of the short-wavelength 
noise that is inherent in the CA dynamics: The deterministic transport dynamics leads to a growth of 
the initially present spatial fluctuations that initially follows the LSA predictions. The CA 
dynamics, by contrast, continually adds spatio-temporal noise at the shortest possible scale 
wavelength, namely on the scale of a single simulation cell. 

In agreement with the LSA the wavelength of the fully developed patterns increases with increasing $A$ or $D$ as illustrated by the stars in \figref{fig:wavelength}. However, at larger $A$ or $D$ values the characteristic wavelengths obtained are larger 
than it is predicted by the LDA. This can be attributed to the extremal dynamics used in the CA 
model. Moreover, the patterns obtained 
from the CA model are much more 'noisy' than their deterministic counterparts. This is evident, for 
instance, from the absence of the higher order 'satellites' in the Fourier patterns of \figref{fig:Daniel_2D_FFT}.

Repeating the simulations at constant external stress leads to results that are virtually identical
with those derived from the quasi-static loading protocol. This is to be expected, since it is in 
the nature of the extremal dynamics that the addition of a constant stress, whatever its magnitude, 
does not change the sequence of events dictated by the extremal rule. As a consequence, the pattern 
wavelengths are stress independent: Whatever the stress level, the fully developed patterns match 
those obtained from the transport model in the limit $\tau_{\rm ext} \to 1$ which represents the 
case of deformation at vanishing rate. This limit actually represents the physically relevant case 
since, in real patterning scenarios, the rate dependent contribution to the flow stress is 
exceedingly small. For illustration, we take typical parameters of Cu where $M_0\approx 2 \times 
10^{4} \, \rm Pa^{-1} \rm s^{-1}$ (see \citet{kubin1992_SM}) and $b=2.54 \times 10^{-10}$ m and assume a 
dislocation density $\rho_0 = 10^{12}$ m$^{-2}$. A typical strain rate of $10^{-3}$ s$^{-1}$ then 
requires a stress of the order of 1 Pa which is about 7 orders of magnitude below the typical 
level of the dislocation interaction stresses, hence, the characteristic deviation of the applied 
stress from the value $\tau_{\rm ext} = 1$ is expected to be negligible.

\section{Conclusions}

Nonlinear simulations of a simple model of dislocation density patterning show that the fully 
developed patterns closely match the predictions derived from a simple linear stability analysis. 
The patterns depend little on the dynamical rules governing dislocation motion: Two different 
dynamic models, one assuming linearly stress dependent viscous dislocation motion and the other an 
extremely jerky cellular automaton evolution with extremal dynamics, produce qualitatively similar results. Also, simulations of the viscous model for different initial conditions show that 
the initial conditions, while having appreciable influence on the transient behavior, are 
practically immaterial to the fully developed pattern. In both viscous and CA models, patterning 
goes along with hardening as evidenced by a reduction in strain rate in the constant-stress viscous 
simulations or by an increase in stress in the CA simulations which use a quasi-static loading 
protocol. The final patterns are essentially governed by a quasi-static balance of the different 
stress contributions entering the model - they depend on a stress balance which makes dislocations 
rest in meta-stable configurations, but NOT on the way the dislocations move between such 
configurations. This provides some hints why dislocation patterns are similar in pure and solute 
hardened fcc metals, or in fcc metals and ionic crystals with KCl structure, where the dislocation 
velocity laws are surely very different. 

Looking at the balance of stresses involved we see three different kinds of stresses, which only
in their mutual interplay can produce the observed patterning: First, we have an external stress 
driving the dislocation system. This is essential: no patterning can take place in the absence of 
plastic flow. Second, we have the stress contributions $\tau_{\rm int}, \tau_{\rm back}$ and 
$\tau_{\rm diff}$ which derive from an energy functional comprising elastic and defect energy 
contributions. These stresses are essential for understanding the pattern morphology and wavelength 
- in particular, the wall-like morphology of the patterns stems from the structure of the elastic 
energy functional and the corresponding stress kernel governing $\tau_{\rm int}$, whereas the 
pattern wavelength depends on the parameters $A$ and $D$ which control the defect energy 
contribution to the energy functional. It is, however, important to note that the internal energy 
related stress contributions alone can {\em not} explain pattern formation: In fact, the patterning 
process depends crucially on a fourth stress contribution which is dissipative in nature, namely the 
friction stress $\tau_{\rm f}$ hence, the present patterning scenario cannot be envisaged as 
'energetically driven'. In fact the basic mechanism leading to instability is simply the fact that, 
in a location of enhanced dislocation density, the friction stress is increased and hence more 
dislocations pile up in the same place. This is precisely the 'dynamic' patterning scenario of 
\citet{Nabarro2000_PMA}, however, with the twist that without accounting for the 'energetic' stress 
contributions it is impossible to understand the pattern wavelength and pattern morphology!

We come thus to the conclusion that much of the past discussion about dislocation patterning may 
have been based upon false dichotomies and misleading analogies. Dislocation patterns are neither 
dynamic dissipative structures nor is their formation driven by energy minimization. Rather, the 
patterns emerge from the attempt of the dislocation system to minimize a coarse grained energy 
functional while driven by an external stress and constantly encumbered by trapping into local, 
'microscopic' energy minima which on the coarse grained scale appear as friction. Past analogies, be 
it with spinodal decomposition or dynamic chemical waves, have in our opinion not been very helpful 
to understanding this interplay. Nevertheless the use of metaphors for conceptualizing dislocation 
patterns has a long (if somewhat murky) tradition and we cannot help coming up with a metaphor of 
our own: We think that the emergence of metastable patterns from an interplay of driving, energy 
minimization and frictional 'shielding' resembles the processes governing the emergence of ripples 
on sand dunes: There, airflow over a sand surface and ensuing saltation provide the external driving 
force, gravity the potential energy that the system tries to minimize, and the screening of fluxes 
by already deposited grains the key mechanism that may lead to instability of a smooth sand surface 
with respect to ripple formation \citep{Kok2012_RPP}. 

\begin{acknowledgments}
M.Z and R.W acknowledge financial support by DFG within the framework of the research unit FOR1650 "`Dislocation
based plasticity"' under grant No Za171/7-1. M.Z. also acknowledges support by the Chinese State Administration of
Foreign Expert Affairs under Grant No MS2016XNJT044. 
IG and PDI  have been supported by the National Research, Development and Innovation Found of Hungary (project no.\ 
NKFIH-K-119561) and the Czech Science Foundation (PDI, project No.\ 15-10821S). PDI is also
supported by the J\'anos Bolyai Scholarship of the Hungarian Academy of Sciences. Finally, we thank 
EC FP7 post-grant Open Access Pilot for financial support on article-processing charges.

\end{acknowledgments}

\bibliography{citepattern}

\end{document}